\documentclass[10pt]{article}
\usepackage[BOE]{express}
\usepackage[]{aas_macros}
\usepackage[]{rotating}
\usepackage{subcaption}
\usepackage{hyperref}
\begin{document}

\title{High--speed dual color fluorescence lifetime endomicroscopy for highly--multiplexed pulmonary diagnostic applications and detection of labeled bacteria}

\author{Ettore Pedretti\authormark{1,7},  Michael G. Tanner\authormark{2,3},  Tushar R. Choudhary\authormark{1,3},  Nikola Krstaji\'c\authormark{3,4,8},  Alicia Megia--Fernandez\authormark{5},  Robert K. Henderson\authormark{4},  Mark Bradley\authormark{5},  Robert R. Thomson\authormark{2,3},  John M. Girkin\authormark{6},  Kev Dhaliwal\authormark{3}  and Paul A. Dalgarno\authormark{1,*}}

\address{\authormark{1} Institute of Biological Chemistry, Biophysics and Bioengineering, School of Engineering and Physical Sciences, Heriot--Watt University, Edinburgh EH14 4AS, UK\\
\authormark{2} Scottish Universities Physics Alliance (SUPA), Institute of Photonics and Quantum Sciences, School of Engineering and Physical Sciences, Heriot--Watt University, Edinburgh EH14 4AS, UK\\
\authormark{3} EPSRC IRC Hub, MRC Centre for Inflammation Research, Queen’s Medical Research Centre, University of Edinburgh, Edinburgh EH16 4TJ, UK\\
\authormark{4} Institute for Integrated Micro and Nano Systems, School of Engineering, University of Edinburgh, Edinburgh EH9 3FF, UK\\
\authormark{5} EaStChem, School of Chemistry, University of Edinburgh, Edinburgh EH9 3FJ, UK\\
\authormark{6} Department of Physics, University of Durham, Durham DH1 3LE, UK\\
\authormark{7} Currently with the Leibniz--Institute f\"ur Astrophysik Potsdam, Deutschland\\
\authormark{8} Currently with the University of Dundee, School of Science and Engineering, Dundee, United Kingdom}

\email{\authormark{*}P.A.Dalgarno@hw.ac.uk} 

\begin{abstract}
We present a dual color laser scanning endomicroscope capable of fluorescence lifetime endomicroscopy at one frame per second (FPS). The scanning system uses a coherent imaging fiber with 30,000 cores. High--speed lifetime imaging is achieved by distributing the signal over an array of 1024 parallel single--photon avalanche photo--diode detectors (SPADs), minimizing detection dead time maximizing the number of photons detected per excitation pulse without photon pile--up to achieve the high frame rate. Dual color fluorescence imaging is achieved by temporally shifting the dual excitation lasers, with respect to each other, to separate the two spectrally distinct fluorescent decays. Combining the temporal encoding, to provide spectral separation, with lifetime measurements  we show a one FPS, multi--channel endomicroscopy platform for clinical applications and diagnosis. We demonstrate the potential of the system by imaging smartprobe--labeled bacteria in ex vivo samples of human lung using lifetime to differentiate bacterial fluorescence from the strong background lung auto--fluorescence which was used to provide structural information.  
\end{abstract}



\bibliographystyle{osajnl}
\bibliography{detectors,photonics,astro,clinical}

\section{\label{intro} Introduction}

Probe--based confocal laser endomicroscopy (pCLE) is a minimally invasive technique used in pulmonary medicine to access the alveolar space\cite{osdoit2007see}. Typically, pCLE takes advantage of the auto--fluorescence from the elastin, collagen, porphyrins, flavin adenine dinucleotide (FAD) and nicotinamide adenine dinucleotide (NADH) present in the lung to provide high contrast, in--situ cellular and sub--cellular--level images of the lung \cite{osdoit2007see}. Such approaches have been hugely successful for pulmonary medicine and there are a number of commercial systems available for clinical use\cite{osdoit2007see}. Examples of clinical applications include invasive pulmonary aspergillosis (IPA) that can be detected by monitoring morphological changes in the alveolar elastin networks \cite{danilevskaya2014case} and the study of macrophage activity\cite{thiberville2009human}. However, despite the success and impact of such systems, relying solely on auto--fluorescence is limiting. Primarily it restricts the ability to identify and distinguish from nearby tissues the specific bacterial colonies that would provide crucial clinical information for diagnosis and treatment.

Currently the most robust solution to identify bacterial colonies is to tag bacteria with fluorescent labels so that during imaging they are spectrally distinct from the inherent background auto--fluorescence. Instruments capable of such multi--wavelength endoscopic imaging\cite{sun2008simultaneous} have been commercially available for some time. We have previously demonstrated a novel multi--channel fluorescence platform designed specifically for endomicroscopy for pulmonary medicine\cite{2016JBO....21d6009K}.  Furthermore, we have also developed selective bacterial fluorescent probes that enable detection and differentiation of bacteria in the human lung \cite{ akram2018, 2017SPIE10041E..0MC}. However, fluorescence endomicroscopy, like all fluorescence imaging modalities, is typically limited to 2 or 3 channels, restricted by spectral mixing that enforces tight bandwidth, filtering and imaging constraints. These challenges are exacerbated in the pulmonary system due to the high intensity green background autofluorescence. 

Fluorescence lifetime imaging (FLIM) is an alternative, well established technique able to discriminate different molecules through fluorescent decay lifetime. The potential of FLIM imaging, using the inherent fluorescence lifetime of the probes over the fluorescence intensity, is well known in the field of optical microscopy, including applications for tissue and clinical use \cite{stringari2011phasor, wallrabe2005imaging,sun2009fluorescence}

Fluorescence species with distinct lifetimes, typically ranging from 1-10~ns, can be distinguished independent of their emission wavelength. When combined with multi--color spectral imaging, FLIM provides an additional selection process, increasing the number of distinct imaging channels. Furthermore, FLIM is generally insensitive to intensity variations or fluorophore density, making it a robust modality for distinguishing subtle changes. However, a major limitation is imaging speed. FLIM typically requires scanning methods to accommodate complex time--tagged single photon counting detection, and significant photon counts to construct meaningful lifetime histograms at each image pixel. 

Fluorescence lifetime imaging endomicroscopy through an imaging fiber bundle has previously been demonstrated, although limited to imaging speeds around 10 seconds per frame. Kennedy et al. \cite{Kennedy:2010:10.1002/jbio.200910065} used a 30,000 core coherent imaging fiber and were able to a measure lifetimes down to 89~ps but with an acquisition time between 10 and 100 seconds per image frame. Bec et al. demonstrated a system \cite{bec2012design}  composed of a single scanning fiber with a moving probe able to record at a 30~Hz rate but requiring 500 images to produce the results shown in the paper, equivalent to 16 seconds acquisition per frame. Yankelevich et al. \cite{yankelevich2014design} used a single scanning fiber, to obtain sub--nanosecond precision lifetime but with an acquisition time of 9.8 seconds per frame. 

In all cases the advantage of FLIM over conventional imaging is primarily dependent on the ability to distinguish separate fluorescence lifetimes, the sensitivity of which is dependent on photon counts, the number of channels in the decay and fitting algorithms. Time--correlated single photon counting (TCSPC) is the most accurate technique for reproducing the true decay times\cite{gratton2003fluorescence}, but is inherently slow, typically delivering a single image frame in 10s of seconds. 

A system that could potentially deliver higher frame rate in endomicroscopy is described by Poland et al\cite{080217e239ee44f88293f086bbdfac25}. Their system is a confocal microscope that uses a "Megaframe" camera (MF32) based on a \(32 \times 32\) SPAD array developed by the University of Edinburgh and STMicroelectronics \cite{richardson200932,Krstajic:15}. The SPAD array has a very low fill factor (1\%) however, a spatial light modulator (SLM) synthesizes \(8 \times 8\) beamlets that are scanned across the sample, before the resulting fluorescence is descanned and directed  onto an \(8 \times 8\) selection of the SPAD array with near 100\% fill factor. The resulting \(256 \times 256\) pixel images were recorded at 10 seconds per frame. However, the speed of acquisition remains limited and the use of the spatial light modulator complicates the optical setup rendering a potential clinical instrument bulky, due to the use of a Ti--Sapphire laser, and in need of complex alignment procedures. 

In this paper we present an alternative approach that uses an entire MF32 SPAD array as a single, parallel SPAD detector, to reduce detection dead time and increase count rate and thus frame rate. We demonstrate 4--channel, one frame per second fluorescence imaging through a combination of dual channel FLIM and dual color, temporally dephased photon counting. We demonstrate the clinical potential of the endomicroscope by imaging {\it ex vivo} human lung with labelled {\it E.coli} bacteria compared to healthy lung.

\begin{figure}[!t]
\begin{center}
\includegraphics[scale=.46]{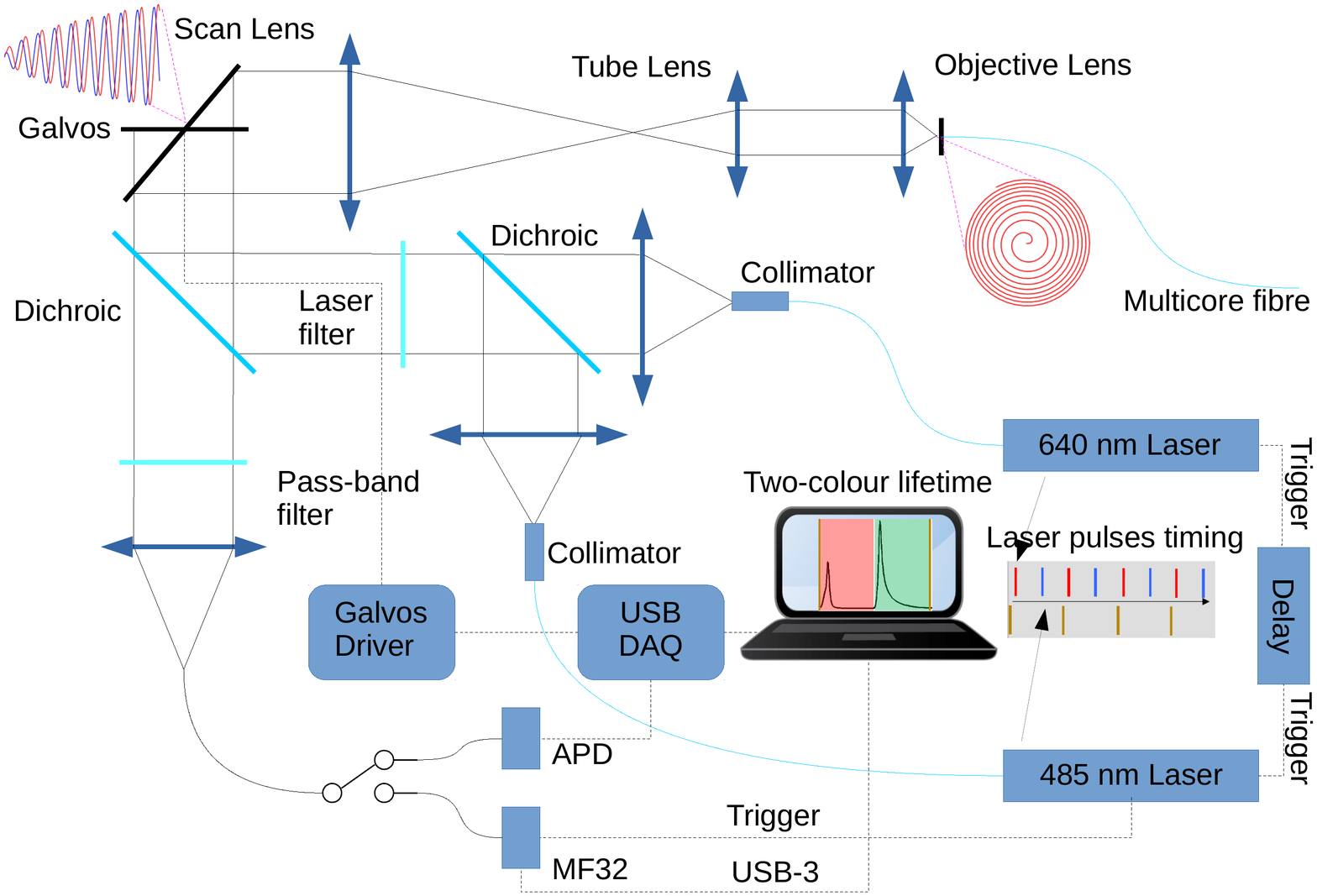}
\end{center}
\caption{\label{LSE} Optics diagram and block diagram of the laser scanning endomicroscope. A USB controlled DAQ generates the scanning waveforms and acquires the analog signal from the APD when in pure intensity scanning mode. The MF32 camera for FLIM data acquisition is controlled through a separate USB interface. Bottom right shows the trigger signal path originating in the 640~nm laser, passing through a delay box, triggering the 485~nm laser and finally triggering the MF32 camera. The emission, dichroic and excitation filters are the XF454 set from Horiba. The schematic shows a raw histogram as acquired from the MF32 camera, showing the red fluorescence decay and the temporally shifted green fluorescent decay}.
\end{figure}

\section{\label{DSC}Description of the instrument}
A schematic of the multiplexed endomicroscope is shown in Figure~\ref{LSE}. Two picosecond--pulsed laser sources (480 and 640~nm, Picoquant GmbH Germany) are driven at a repetition rate of 10~MHz. The trigger output of the 640~nm laser source drives the 480~nm source through an analog and programmable delay box, additionally triggering the MF32 camera to start the TCSPC acquisition. The two spectrally distinct fluorescence signals from green and red excitation are thus separated in time through this temporal offsetting of the two excitation lasers. The two laser sources are fiber coupled into two collimators, whose output are combined via a dichroic (FF506-DI03-25X36, Semrock USA) before passing through a 10 nm band--pass laser filter to clean the laser emission. The resulting beam is then scanned using a pair of close coupled galvos (ThorLabs Inc, USA). 

The resulting scanned beam is conveyed through a CLS-SL visible scan lens (400--750~nm, Thorlabs, EFL 70~mm) and an ITL200 infinity--corrected tube lens (Thorlabs, f 200~mm), to a Nikon plan fluorite imaging objective, \(\times\)10, 0.3 NA, that focuses the excitation light onto a multi--core fiber. The fiber has ~30,000 cores, with sizes varying from 5-10 microns (Fujikura, Japan). The laser scan system is achromatic for the two excitation wavelengths but the imaging fiber is not. To mitigate this the system is aligned such that coupling is equally compromised for both excitation wavelengths.  

The fluorescence is collected back from the distal end of the fiber bundle and the returning light is collected by the objective and passes back through the optical system to be descanned, before being transmitted by the dichroic filter and passing through a pass--band filter to ensure no laser light enters the M43L01, 105~\(\mu\)m  0.22 NA, multimode fibre that acts also as an optical pinhole before the detectors. The fluorescence signal, the output of the multimode fiber, can be switched to either an APD detector for 10~FPS fluorescence intensity images, or defocused to fill the MF32 camera for 1~FPS FLIM imaging.

\begin{figure}[!t]
\begin{center}
\includegraphics[angle=0, scale=.35, trim={2.0cm 1.0cm 1.0cm 1.0cm}]{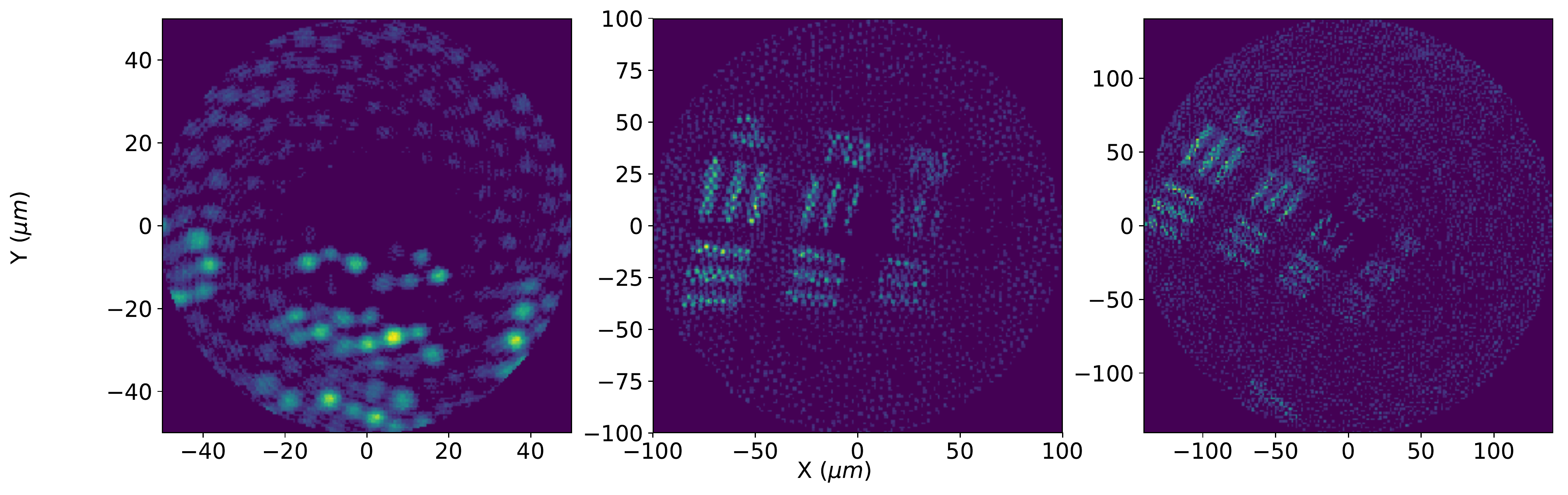}
\end{center}
\caption{\label{coreMagMovie} Fluorescent image sequence acquired in non time--resolved mode of a 1951 USAF resolution test chart. The sequence starts left with the highest resolution and ends right with lower resolution, zooming out of group 7, element 2. The single cores of the imaging fibre (5 \(\mu m\)) are visible in the highest resolution image.}
\end{figure}

The non--resonant scanning galvos are driven to achieve one frame per second imaging rate in a raster scan mode. However, by using a spiral pattern we were able to obtain frame rates in excess of 10~FPS in fluorescence imaging mode. When in the spiral scanning mode the scan matches the circular fiber bundle closely. An important advantage of using non--resonant galvos is that they allow us to freely control scan rates, patterns and ranges (e.g. Figure~\ref{coreMagMovie}). For example, the architecture described here is appropriate for performing wide field of view imaging, for focusing on specific targeted areas of tissue or for addressing single fibre cores.

To demonstrate the basic performance of the system, Figure~\ref{coreMagMovie} shows a 1951 USAF resolution test chart coated on the underside with an ultra--high density of fluorescent microspheres, imaged in fluorescent mode using the APD. The sequence starts with a high--resolution image of group 7, element 2 of the chart, where the single cores of the fiber are visible and zooms out to a larger field of view. Optical resolution is limited by both core spacing (around 5 \(\mu m\)) and by core to core coupling\cite{Perperidis:17}. 

In FLIM modality the system uses the MegaFrame  \(32 \times 32\) SPAD array as a single--pixel detector, rendering the detection from the 1024 individual SPAD detectors parallel. In the low photon detection limit employed here, incident photons stochastically impact on each pixel in the array and the signals from each of the 1024 SPADs detectors are then binned in a single temporal histogram of the photon arrival time for each single frame from the SPAD array producing the signal of a single pixel in the final image. Crucially this parallelization increases the maximum detectable number of photons per unit time enabling more efficient use of the fluorescently excited photons with a high counting rate. It has previously been demonstrated \cite{ceccarelli2017development}, that the use of a \(8 \times 8\) SPAD array, as a single--pixel detector, showed a factor 64 increase in photon rate.

In this system operation, photon counts per detector are below the maximum detector readout rate of hundreds of kHz \cite{Krstajic:15} (total count rate increased to hundreds of MHz by detector parallelization over the whole array), while laser repetition rate is 10~MHz. If an imagined (the authors are not aware of such a device) single detector capable of 100~MHz photon count rates were used (equivalent to the maximum count rates possible here), and operated near these rates then photon pileup \cite{becker2005advanced} would be overwhelming. Even at two orders of magnitude lower rates (\~MHz) pileup would cause perturbation of the observed photon timing statistics. This would lead to shortened observed fluorescence lifetimes, and variations of this artifact with varying numbers of photon counts. Meanwhile, at the 100~MHz maximum possible count rates in our system each detector is operating at only 1.0\% of the laser repetition rate, a regime recognized to be safe from photon pileup \cite{becker2005advanced}.

The red and green photons reaching the detectors of the MF32 camera are separated in time due to the delay introduced in triggering the blue laser with respect to the red laser, this delay being much longer than any of the fluorescence lifetimes, as shown in Figure~\ref{LSE} . Counts in one color do not have an effect on the other channel (due to photon pileup) in the noted regime where counts are low per detector compared to laser repetition rate. Only in such a parallel detection scheme is high speed FLIM not in danger of photon pileup artifacts, and temporal color multiplexing becomes viable as described.

Histogramming the photon counts across all parallel pixels of our detector also combines the dark counts present. Statistical noise in a measurement is dependent on the square root of the counts. As such, parellization which simultaneously increases both photon counts and dark counts results in less noisy counting statistics than those of a count rate limited system.

For fast lifetime imaging we use a center of mass (C--O--M) algorithm \cite{080217e239ee44f88293f086bbdfac25} for real time calculation of the exponential lifetime for both the green (485 nm excitation) and red (640 nm excitation) fluorescence.). As in standard beam scanned FLIM, the photon counting occurs for a fixed dwell time for each scanned point, building up a complete image during the scan. Raw data was saved as flexible image transport system (FITS) format \cite{1981A&AS...44..363W} as a sequence of \(32 \times 32\) images from the MF32 with the associated scanner position enabling multiple alternative methods suitable for  processing to obtain fully fitted TCSPC FLIM images. The system scan speed is limited to one frame per second by the SPAD array to FPGA transfer speed.

\section{\label{RES} Results}

We demonstrate the potential of the instrument by first using an example system of a multi--core fiber with 19 widely spaced cores with the distal end loaded with fluorescent microspheres, then utilizing a 30,000--core imaging fiber to view microspheres on a slide, and then viewing {\it ex vivo} human lung--tissue to show the detection of bacteria against a strong spectrally similar fluorescence background using lifetime as the contrast mechanism. Images were obtained in 10--FPS fluorescence mode and in 1--FPS FLIM mode, in single and dual color modalities. 

\subsection{\label{beadsFLIM} FLIM capability test with fluorescent microspheres}
\begin{figure}[t]
\centering
\includegraphics[width=1\linewidth]{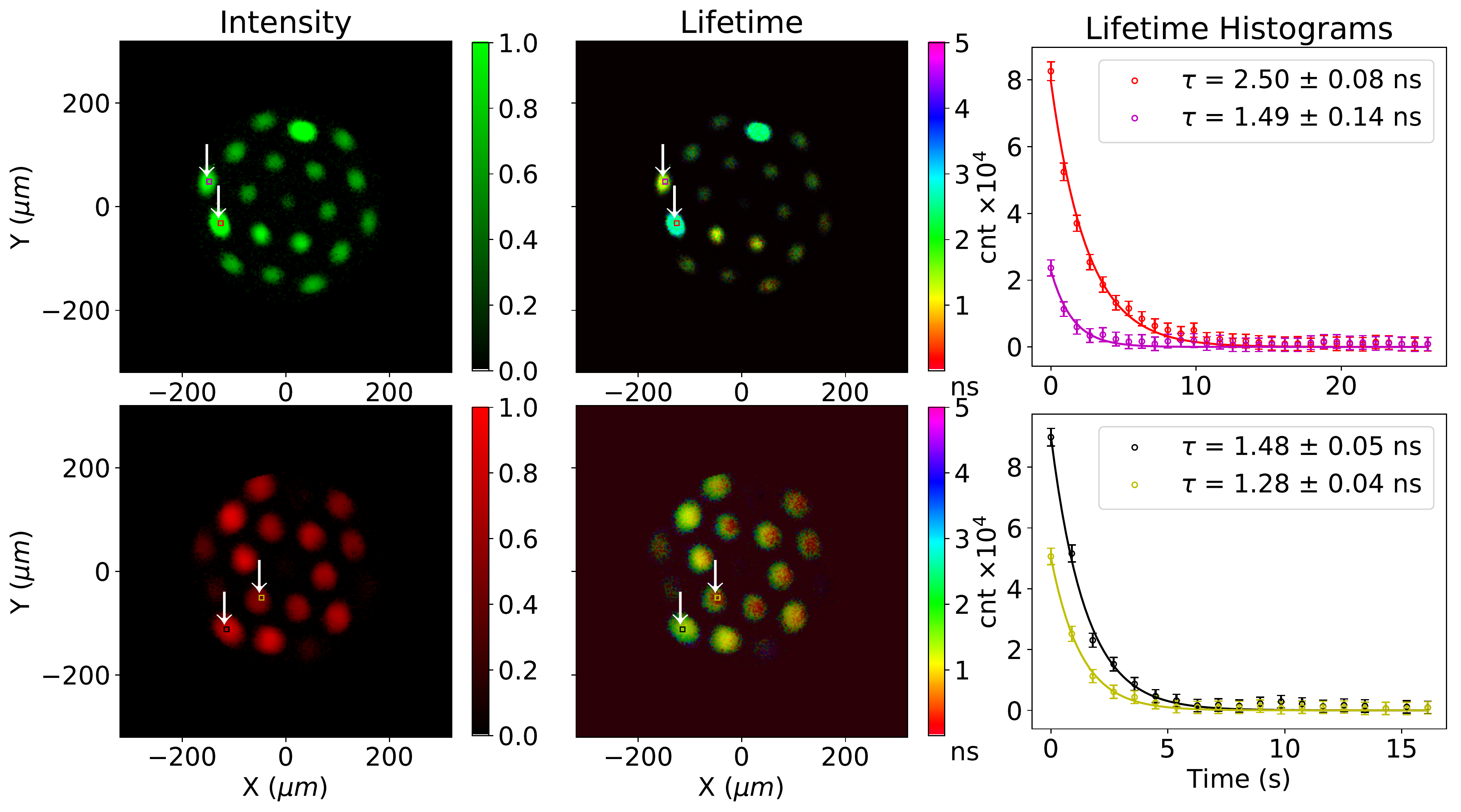}
\caption{\label{histBeads} 
A multi--core fiber loaded with four different fluorescent microspheres acquired at 1 FPS. Top shows the green channel and bottom the red channel. The left hand column shows a fluorescence image, the center column a lifetime image. The plots show the lifetime measured from two distinct areas of \(9 \times 9\) elements in the figures, marked by red and purple squares on the top row and black and yellow squares on the bottom row. The plots  and the uncertainties were obtained using a single exponential fit with the scipy.optimize.leastsq module \cite{jones2014scipy}. The FLIM images use the center--of--mass method \cite{080217e239ee44f88293f086bbdfac25}.}
\label{fig:spiral}
\end{figure}

To demonstrate the basic operation of the system a loaded fiber was used in place of the normal imaging fiber. Four types of microspheres were loaded on to the distal end of a 19--core multimode fiber with  \(10~\mu m\) cores. Pits \(10~\mu m\) in diameter were formed by selectively etching the cores using hydrofluoric acid, into which microspheres self--locate. This configuration is used for sensing pH and other physical parameters, using procedures described in \cite{choudhary2017multiplexed,choudhary2018SR}. Two types of microspheres had 
``red'' emission, two with ``green'' emission and each with different lifetimes. Specifically we used \(10~\mu m\) silica microspheres  covalently bound to Fluorescein (green), TAMRA (orange), NBD (yellow) and CY5 (red) dyes. The Results are shown in Figure~\ref{histBeads}.

Figure~\ref{histBeads} shows in the top panels the green channel and the bottom panels the red channel. On the left side intensity images are shown while in the center lifetime images are shown. The right side of Figure~\ref{histBeads} shows the average histogram of the photon arrival time from the two red and purple square areas drawn on the intensity and FLIM images. 

We find that the resulting histogram obtained for each scanned point is more than adequate to produce FLIM images at the rate of one FPS. The histograms of Figure~\ref{histBeads} each show two exponential decays for the same wavelength of emission from the two lifetimes for the microspheres present. The green channel in the top plots shows two distinct lifetime species, 1.51~ns and 2.56~ns. The red channel also shows two distinct lifetime species, 1.47~ns and 1.29~ns. Distinction of these species is subjective in the intensity only images, but quantifiable in the lifetime images. Thus our system can easily distinguish 4 distinct species, through both spectral and temporal identification. Temporal accuracy is limited by the laser pulse width, which is approximately 100~ps\cite{richardson200932}. This result demonstrates the ability to resolve signals from four different probes through multi--spectral FLIM at 1 FPS.

The lifetime images were weighted by the fluorescent images so that low signal--to--noise (SNR) pixels are not shown. Weighting was done by modulating the alpha channel, which controls the transparency of the image, of each FLIM image with the fluorescent image. In this way low SNR pixels become transparent and disappear from the image, but the weighting does not impinge on the color scale for the lifetimes as traditional intensity weighting would.

\begin{figure}[!p]
\begin{center}
\includegraphics[angle=0, scale=0.29]{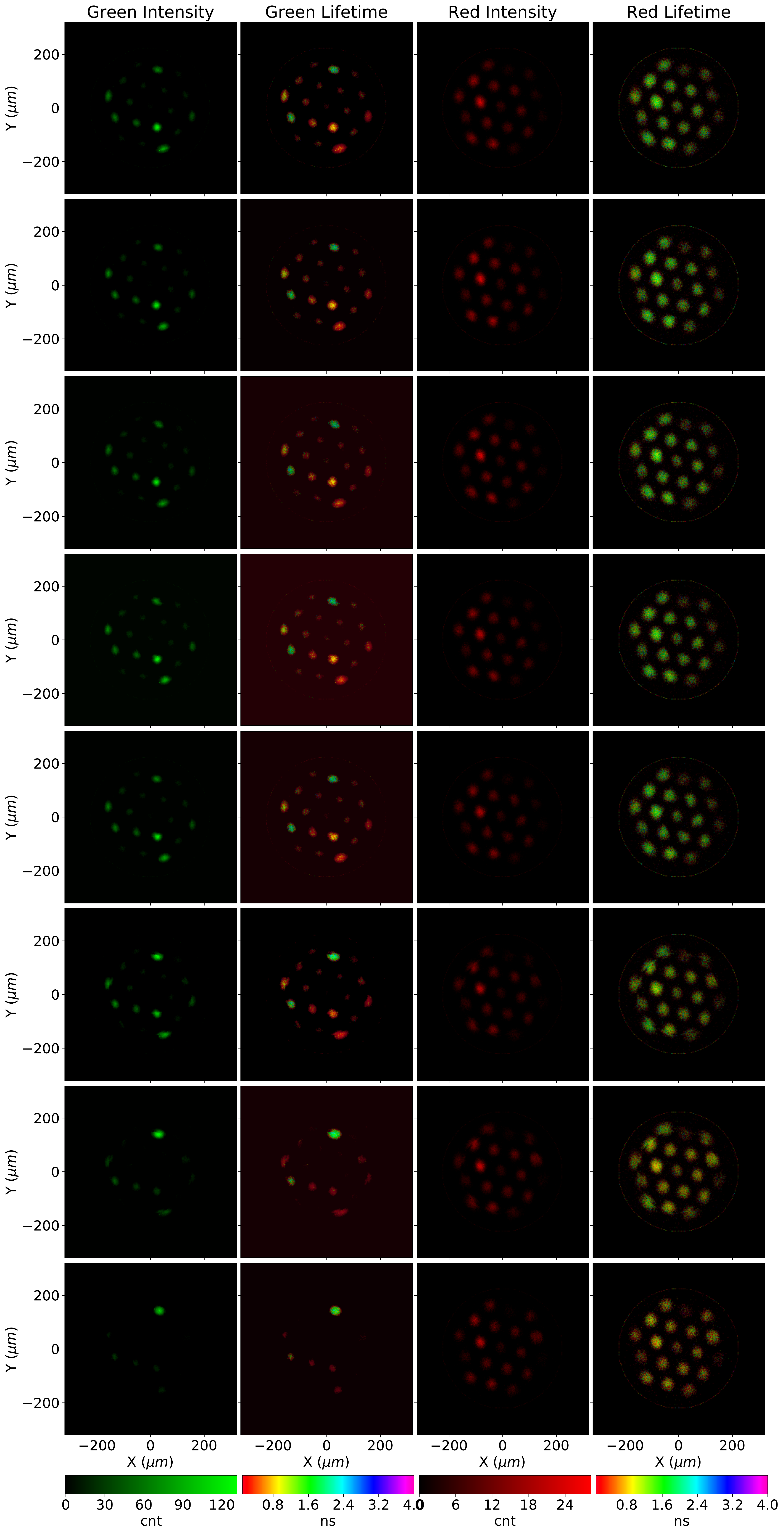}
\end{center}
\caption{\label{movieLT} Sequence of eight images extracted from a ten--frames, 1--FPS FLIM movie. Microspheres covalently bound to Fluorescein (green), TAMRA (orange), NBD (yellow) and CY5 (red) dyes were loaded onto a mult--core fiber. The fiber was placed in water during recording, causing most microspheres to show a decrease in lifetime.}
\end{figure}

To demonstrate high speed FLIM we include frames from dynamic measurements in the Appendix. Figure~\ref{movie} shows  8 extracted frames from a 1--FPS movie, imaged under similar conditions as described for Figure~\ref{histBeads} but utilizing the imaging fibre. The sample consisted of mixed red and green fluorescent microspheres of the same type as described in the previous experiments. They were deposited on a microscope slide as a liquid solution and let dry. The 30,000 core imaging fibre was put nearly in contact with the slide and shifted manually using a translation stage. The microspheres in Figure~\ref{movie}  are seen shifting from right to left. Each row represents a single temporal frame. Frame one is at the top. The columns from the left show a green fluorescent image, a green lifetime image, a red fluorescent image and a red lifetime image. Figure~\ref{movieLT} shows an additional movie. We used the same sample as in Figure~\ref{histBeads}. This time the fibre probe with fluorescent microspheres were placed in water during the capture. The red microspheres show a decrease in lifetime from 1.8~ns to about 1.2~ns after water was added. The green microspheres show a decrease of lifetime from 0.9~ns to about 0.1~ns while additional green microspheres show a decrease from 2.4~ns to 1.6~ns.

\begin{figure}[!t]
\begin{center}
\includegraphics[scale=.18]{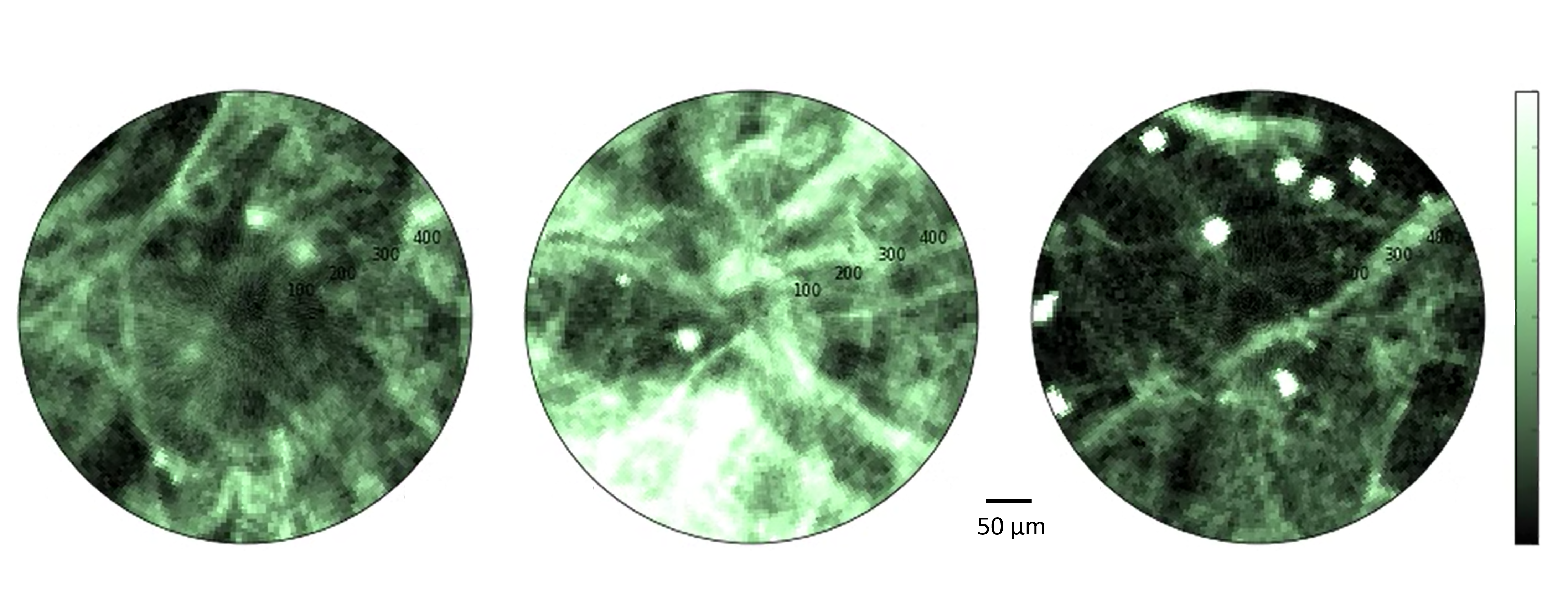}
\end{center}
\caption{\label{lungPlusMonicites} Above are three frames extracted from a 10~FPS movie from {\it ex vivo} lung tissue taken in in non time--resolved modality using the APD detector. Some monocyte cells labeled with Calcein are visible in the foreground. The image intensity represents the voltage output of the APD (in response to fluorescence intensity) on the indicated color scale. The field of view is 500 $\mu$m. Parameters were kept constant throughout the image frames giving good contrast of lung tissue, however some saturation of brighter features is therefore present.}
\end{figure}

\subsection{\label{BAC} {\it Ex vivo} human lung tissue imaging with labeled bacteria}
\begin{figure}[t]
  \centering
  \includegraphics[width=1.0\linewidth]{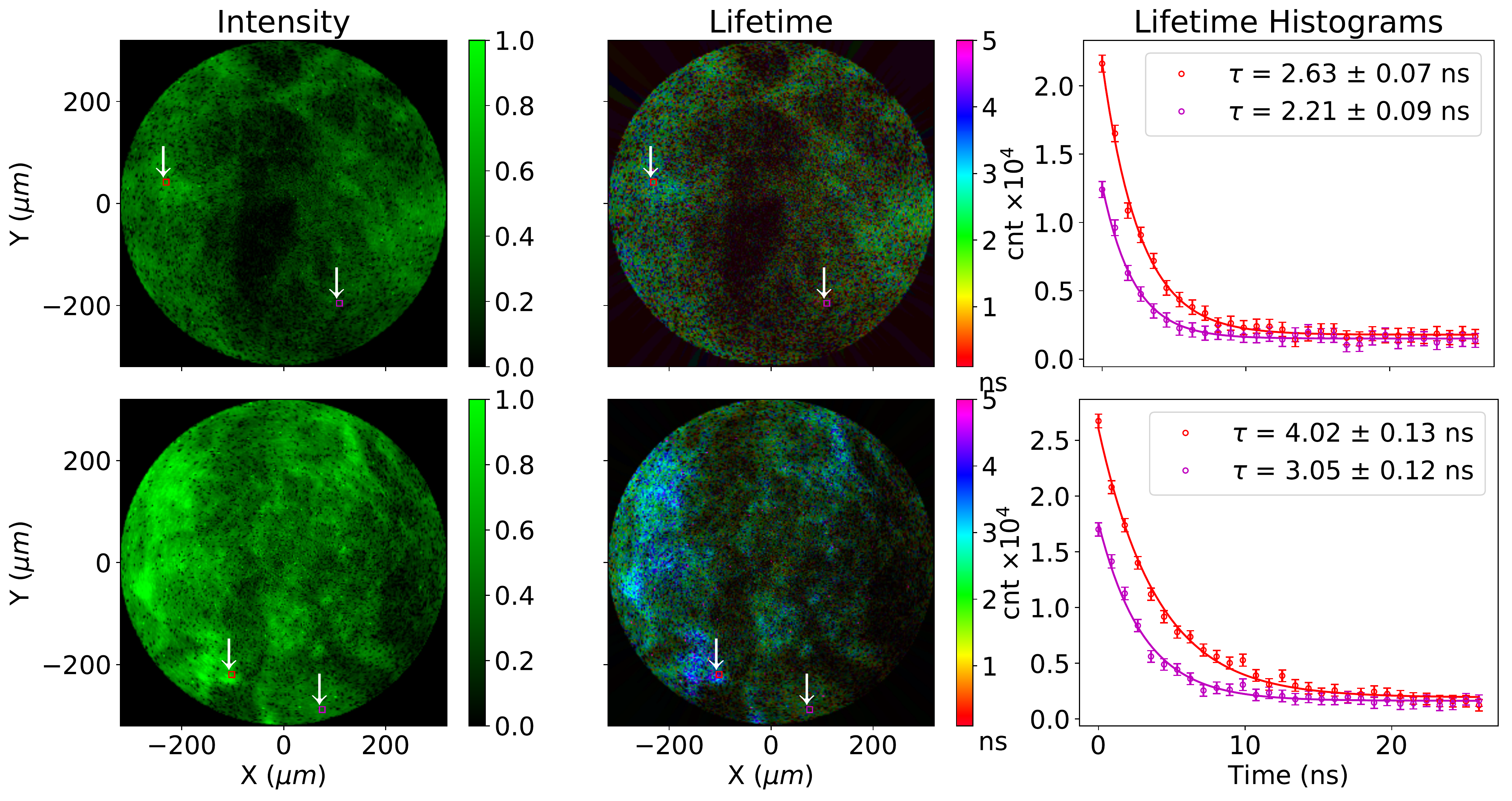}
\caption{\label{lung-Bac2} {\it Ex vivo} human lung tissue. The top row shows human lung and bottom row shows human lung plus {\it E.coli} and NBD--PMX smartprobe.  The rightmost plots show the lifetime measured from two distinct \(9 \times 9\) element areas in the figures, represented by a red and a purple squares.}
\label{fig:Lung}
\end{figure} 
To demonstrate the clinical significance of the instrument experiments were undertaken on {\it ex vivo} human lung samples artificially infected with labeled bacteria. Human lung samples were obtained from patients undergoing surgical resection for lung carcinoma. All lung sections used in this study were sections of normal healthy lung away from the cancerous growth. The study was approved by the Regional Ethics Committee (REC), NHS Lothian (reference 15/ES/0094), and with informed consent of the patients. Human lung tissue samples were dissected into thin, small sections (4 mm by 4 mm) and placed onto a 96--well tissue culture plate for imaging.

Figure~\ref{lungPlusMonicites} shows an extract from a 10~FPS intensity only movie of {\it ex vivo} lung populated with live fluorescent monocyte cells of approximately(9 \(\mu m\)) and labeled with Calcein. (see supplementary Movie S1 for full video) The movie was acquired using the APD single pixel detector. As discussed previously, human lung is naturally autofluorescent in the green mainly due to the presence of elastin and collagen in the connective tissue therefore Calcein provides contrast by making the monocyte cells brighter. Functioning in this conventional modality the system demonstrates the observation of lung structures at 10~FPS through fibre, but it is clear that weak signals from fluorophores could be obscured by the lung autofluorescence.

%
\subsection{\label{lungBAC2FLIM} Human lung with NBD--PMX bacterial imaging smartprobe}
\begin{figure}[!t]
\begin{center}
\includegraphics[angle=0, scale=0.25]{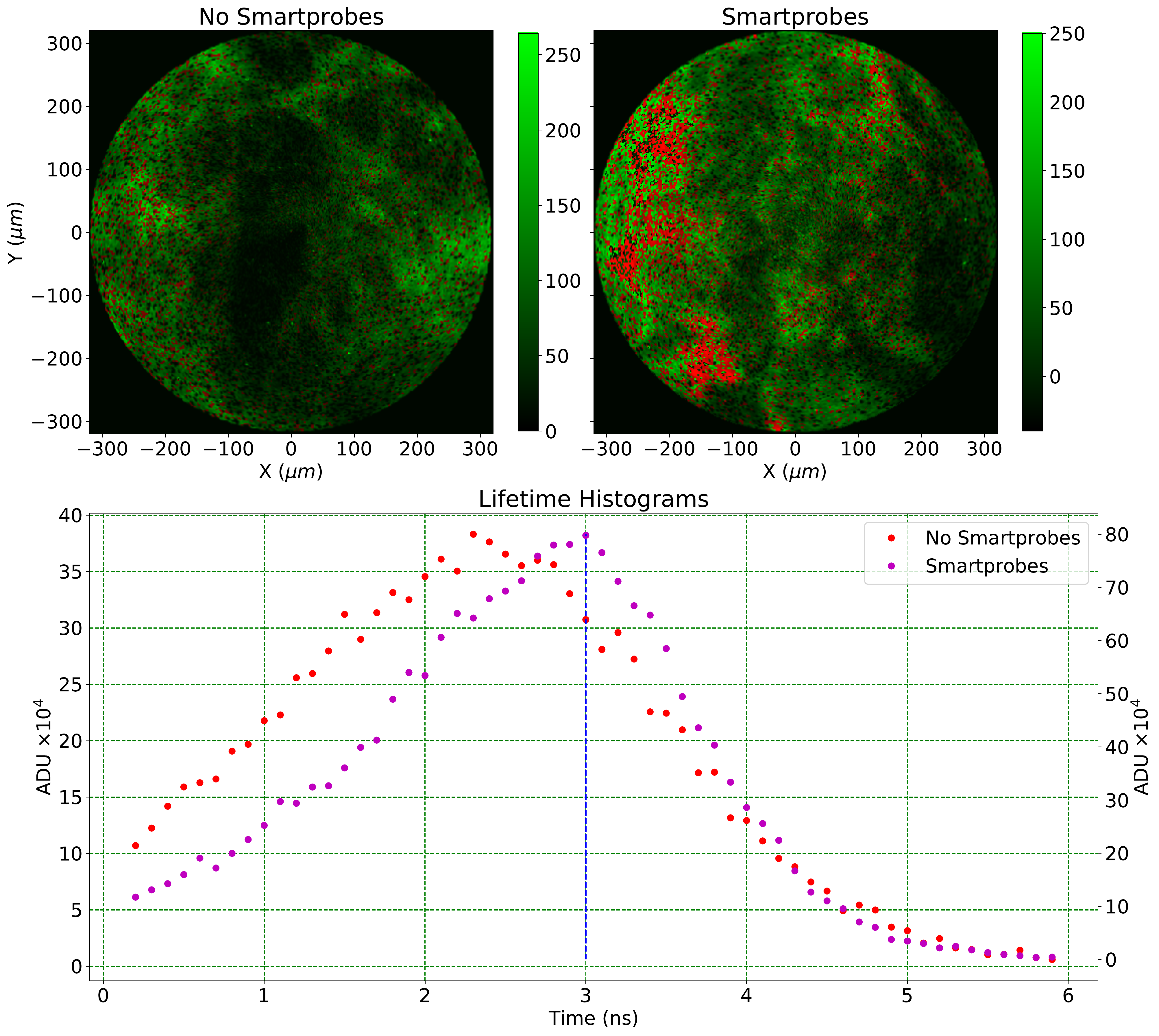}
\end{center}
\caption{\label{Thresholding} Top left is {\it ex vivo} lung and top right is {\it ex vivo} lung with NBD--PMX smartprobe labeled bacteria. The bottom panel shows two histograms of the lifetime distribution from the whole images. The red histogram is from the left image without smartprobes and the purple histogram from the right image with smartprobes. A threshold of 3~ns was applied to the FLIM data to identify regions where the smartprobes were active. The red dots on the right image represent the region where lifetime was larger than 3~ns and counts were larger than 50.}
\end{figure}

Overnight cultures of {\it E.Coli} grown in LB broth (250 rpm, 37 $^{\circ}$C) were washed (x3) in phosphate buffered saline (PBS, Gibco) and then adjusted to an OD\(_{595}\) of 2. The bacteria were labeled with a NBD--PMX smartprobe \cite{2017SPIE10041E..0MC,akram2018} at \(5 \mu M\) concentration and were washed in PBS to remove any excess dye after approximately 1 minute. For imaging, pre--labelled {\it E. Coli} (\(100 \mu L\)) were added to a section of lung tissue in a 96--well tissue culture plate and imaged immediately.

Figure~\ref{lung-Bac2} (top) shows {\it ex vivo} human lung tissue imaged in lifetime mode with the MF32 and C--O--M analysis described previously.  Healthy lung tissue (top) and that artificially infected with {\it E.coli} labelled with a NBD--PMX smartprobe (bottom) are shown. The left images show intensity only, taken from summing photon counts from the FLIM data, and the right images the intensity weighted FLIM images. All images were recorded at a rate of 1 FPS, but here 10 static frames have been combined to provide a clearer image. The rightmost plots show the average histogram of the photon arrival times from the two red and purple squares on the fluorescent and FLIM images. The intensity only images show a characteristic increase in intensity for the infected tissue, but identification of the smart probe labeled bacteria is challenging over the spectrally overlapping autofluorescence. In lifetime mode the distinction between bacteria (with lifetime 4.0~ns) and autofluorescence (lifetime 3.0~ns) is much clearer. Thus lifetime here acts as an excellent contrast mechanism\cite{2017SPIE10041E..0MC}.

To further exemplify this distinction, Figure~\ref{Thresholding} shows the same data from Figure~\ref {lung-Bac2} with temporal thresholding applied using the histograms from the lifetime of the two images to determine the threshold. The top panel reproduces the FLIM image with and without smartprobe labeled bacteria, the bottom panel shows a binned histogram of each image. A 3~ns threshold, corresponding to the peak of lifetimes in the smart probe labeled data, was applied to the top images and the resulting pixels with lifetime above this threshold are then colored red and shown in both images of Figure~\ref{Thresholding} but only notable in the right figure when smart probe labeled bacteria were present. With this example of thresholding, tuned to the smartprobe lifetime, the smartprobes become immediately visible in the FLIM image while being hidden in the fluorescent image.

This result demonstrates the clear potential for distinguishing fluorescently labelled bacteria from the intrinsic auto--fluorescence of lung tissue, based on lifetime through an imaging fiber at one frame per second with simple, real time enabled processing algorithms. This method is appropriate for {\it in vivo} clinical use, where it is not normally possible to spectrally separate the smartprobe from lung autofluorescence\cite{akram2018}.

\section{Conclusions}
We have developed and built a dual-color, FLIM enabled laser scanning endomicroscope capable of fluorescence lifetime imaging endomicroscopy at one frame per second. The system uses a Megaframe \(32 \times 32\) SPAD array detector as a single--pixel detector, with all 1024 detectors contributing to the buildup of a histogram per scanned image element.

A center--of--mass algorithm with background subtraction was then used to determine the lifetime. We obtained FLIM images at a rate of one frame per second and show standard fluorescence imaging at a frame rate of 10~FPS with non--resonant galvos, using a spiral scan pattern rather than the more conventional raster scan. This enables a more flexible endomicroscopy system, capable of selecting regions of interest with controllable zoom, scan range and scan rate.

We have shown that the endomicroscope is capable of dual--colour imaging by separating the green and red channel in time using our high speed fast recovery detector. We have also shown that each single color channel can discriminate microspheres of different lifetime. We detected {\it E.coli} bacteria in {\it ex vivo} human lung using a fluorescent NBD--PMX bacterial imaging smartprobe\cite{akram2018} that provides longer lifetime in FLIM imaging (4.02\(\pm\)0.13~ns) from a lung backgroung of 2 to 3-ns. This demonstrates that using integrated SPAD arrays and lifetime imaging it is possible to use fluorescence lifetime as a contrast mechanism to help remove background auto-fluorescence in human tissue. The method is suitable for clinical use with a current imaging frame rate of 1 frame per second currently limited by the data transfer speeds.

\section*{Funding}
We thank the Engineering and Physical Sciences Research Council (EPSRC, United Kingdom) Interdisciplinary Research Collaboration grant EP/K03197X/1 for funding this work.

\section*{Acknowledgments}
This research has made use of NASA's Astrophysics Data System Bibliographic Services. This research made use of IPython \cite{PER-GRA:2007}; Matplotlib, a Python library for publication quality graphics \cite{Hunter:2007}; SciPy \cite{jones2014scipy}, a Python library for scientific computing; Cython \cite{behnel2011cython}, a Python library for C/C++ code wrapping.

\section*{Disclosures}
 The authors declare that there are no conflicts of interest related to this article.
 
\section{Appendix}

\begin{figure}[!p]
\begin{center}
\includegraphics[angle=0, scale=0.29]{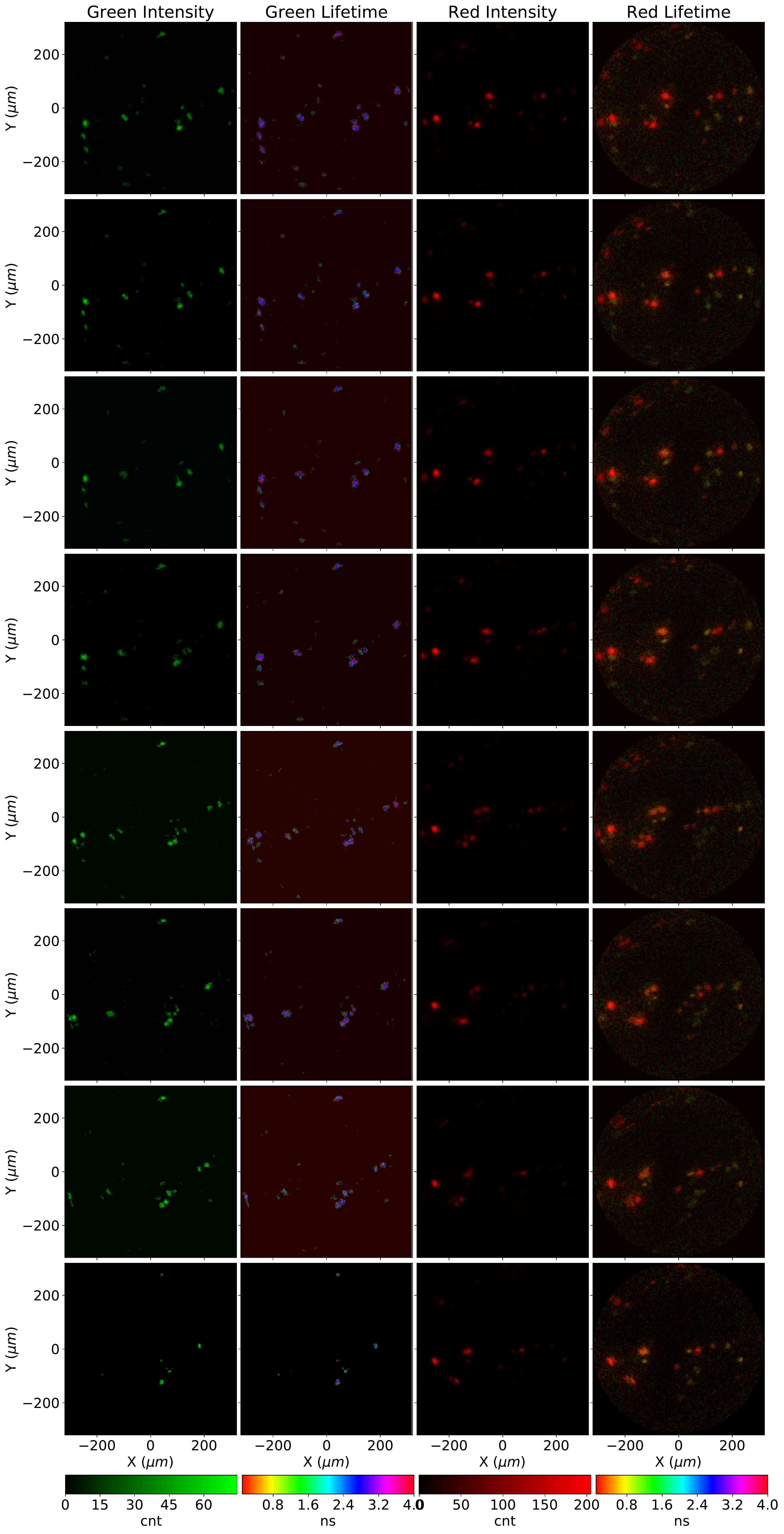}
\end{center}
\caption{\label{movie} Sequence of eight images extracted from a ten--frames, 1--FPS FLIM movie. Microspheres  covalently bound to Fluorescein (green), TAMRA (orange), NBD (yellow) and CY5 (red) dyes were deposited in a microscope slide. The imaging fiber was mounted on a XYZ translation stage and moved a few \(\mu m\) across the surface of the slide.}
\end{figure}

\end{document}